\documentclass{article}
\usepackage{amssymb}

\input{tcilatex}
\begin{document}

\title{On Time(s) in Quantum Mechanics and Quantum Gravity: the demise of
Pauli's objection and an integrating new perspective}
\author{M. Bauer* and C.A. Aguill\'{o}n** \\
%EndAName
*Instituto de F\'{\i}sica, **Instituto de Ciencias Nucleares,\\
Universidad Nacional Aut\'{o}noma de M\'{e}xico\\
Ciudad Universitaria, CP 04510, M\'{e}xico, CDMX, MEXICO\\
e-mail: bauer@fisica.unam.mx}
\maketitle

\begin{abstract}
Dirac's canonical quantization applied to closed systems leads to static
equations, the Wheeler-deWitt equation in Quantum Gravity (QG) and the time
independent Schr\"{o}dinger equation in Quantum Mechanics (QM). How to
restore time and what time it is, constitutes the Problem of Time(s), still
unresolved after a century of the development of QM and QG. A prominent role
all along has been played by Pauli's objection to the existence of a time
operator, relegating time to be a parameter and leading to statements such
as: "time in QM and time in QG are mutually incompatible notions".

Integrating developments of a new two times perspective\ are:

a) the emergence of the time dependent Schr\"{o}dinger equation from the
entanglement of a microscopic system with its classical environment, that
accords to the microscopic system a time evolution description, where $t$ is
the laboratory time;

b) the canonical quantization of Special Relativity (SR) correctly shown to
yield both the Dirac Hamiltonian and a self adjoint "time" operator,
restoring to position and time the equivalent footing accorded to energy and
momentum in Relativistic Quantum Mechanics (RQM). It introduces an intrinsic
time property $\tau $ associated with the mass of the system, and an
additional orthogonal basis to the usual configuration, momentum and energy
basis. Being a generator of continuum momentum displacements and
consequently of energy, it invalidates Pauli's objection to the existence of
a time operator. It furthermore complies with the requirements to condition
the other observables in the conditional interpretation of QG. As Pauli's
objection figures explicitly or implicitly in most current developments of
QM and QG, its invalidation opens to research the effect of this new two
times perspective on such developments.

keywords: time operator; entanglement; Pauli's objection; Quantum Mecanics;
Quantum Gravity
\end{abstract}

- \textbf{"If you are receptive and humble, mathematics will lead you by the
hand"}

- "\textbf{One must be prepared to follow up the consequences of theory, and
feel that one just has to accept the consequences no matter where they lead"}

\ \ \ \ \ \ \ \ \ \ \ \ \ \ \ \ \ \ \textbf{\ }\ \ \ \ \ \ \ \ \ \ \ \ \ \ \
\ \ \ \ \ \ \ \ \ \ \ \ \ \ \ \ \ \ \ \ \ \ \ \ \ \ \ \ \ \ \ \ \ \ \ \ \ \
\ \ \textbf{P. A. M. Dirac }

\section{Introduction}

Dirac's canonical quantization of the Hamilton-Jacobi formulation applied to
closed systems leads to static equations, the Wheeler-deWitt equation in
Quantum Gravity (QG)\cite{DeWitt,Wheeler} and the time independent Schr\"{o}%
dinger equation (TISE) in Quantum Mechanics (QM)\cite{Schroedinger}. The
time dependent Schr\"{o}dinger equation (TDSE), as presented originally by
Schr\"{o}dinger in 1926\cite{Schrodinger2}, is the result of an educated
replication of Hamilton's optical-mechanical analogy rather than a formal
derivation.\ 

How to introduce time in QM, an object of research for decades\cite%
{Muga,Muga2}, needs to deal with Pauli's objection to the existence of a
time operator canonically conjugate to energy in QM. To quote\cite{Pauli}:"
In the older literature on quantum mechanics, we often find the operator
equation $[H,T]=-ihI$.\textbf{.....} It is generally not possible, however,
to construct a Hermitian operator (e.g. as function of p and q) which
satisfies this equation. This is so because, from the C.R. written above, it
follows that $H$ possesses continuously all eigenvalues from -$\infty $ to +$%
\infty $ , whereas on the other hand, discrete eigenvalues of $H$ can be
present. \textit{We, therefore, conclude that the introduction of an
operator }$\mathit{T}$ \textit{is basically forbidden and the time t must
necessarily be considered as an ordinary number ("c-number') in Quantum
Mechanics". }As a consequence\ QM fails to treat time and space on an
equivalent footing accorded by Special Relativity (SR). In QM time appears
as a parameter, not as a dynamical variable represented by a self adjoint
operator. It is a c-number, following Dirac's designation\cite{Dirac}
resulting in the extended discussion of the existence and meaning of a time
operator and of a time energy uncertainty relation\cite%
{Muga,Bush,Bauer6,Bauer}. Most notable is the extension of the von Newman
definition of observables to Hermitian but not self adjoint operators (e.g., 
$\hat{T}=\frac{m}{2}\left\{ \frac{1}{\hat{p}}\hat{x}+\hat{x}\frac{1}{\hat{p}}%
\right\} $ of Bohm and Aharonov\cite{Bohm}, based on the non relativistic
Hamiltonian $H=\frac{p^{2}}{2m})$, and the concept of positive operator
valued measurement (POVM)\cite{Olkhovsky,Recami}. One can also note that
within the time quantities considered in the litterature one finds
instantaneous values and intervals, e.g., parametric (clock) time, tunneling
times, decay times, dwell times, delay times, arrival times or jump times.
To quote the introduction in Ref.5: \textquotedblleft In fact, the standard
recipe to link the observables and the formalism does not seem to apply, at
least in an obvious manner, to time observables\textquotedblright . This is
the Problem of Time in QM.

Quantum Field Theory (QFT) has often been considered a solution as position
is no longer associated to an operator and $x$ and $t$ are joined as space
time field coordinates. In this respect it is important, following Hilgevoord%
\cite{Hilgevoord}, to clearly avoid the confusion between the space
coordinates $(x,y,z)$ of a system of reference and the quantum mechanical
position operator $\mathbf{\hat{r}=(}\hat{x},\hat{y},\hat{z})$\ whose
expectation value gives the time evolution of the position of a system
described by a certain state vector $\mid \Psi (t)>$. To quote: "If by space
and time one understands the coordinates of a given space and time
background, none of these coordinates are operators in quantum mechanics.
If, on the other hand, one thinks of position and time as dynamical
variables connected with specific physical systems situated in spacetime,
the representation of such variables by quantum mechanical operators is
possible". A distinction should be made between the time coordinate in $%
(x,y,z,t)$ and a "time operator $\hat{T}$ acting on the system state vector.
The space coordinates $(x,y,z)$ together with the parameter $t$ constitute
the background spacetime SR framework of an inertial observer, whereas the
operators $\mathbf{\hat{r}=(}\hat{x},\hat{y},\hat{z})$ and, if it exists, $%
\hat{T}$ , are to be considered as, say, "path" dynamical space and time
operators, in the same way as energy and momentum.

Quantum Gravity (QG), the quantization of General Relativity (GR), is still
an unsolved problem in physics. One of the main difficulties arises from
accepting that time in quantum mechanics is a parameter, whereas in GR
matter determines the structure of spacetime with time and space acquiring a
dynamical nature. Thus "time" in QM and "time" in GR are seen as mutually
incompatible notions. This is the Problem of Time in QG\cite%
{Anderson,Isham,Kushar2,Butterfield,Kiefer,Ashtekar,Smolin,Rovelli}.

In the present paper a clarifying and integrating picture that recognises
two different times is shown to arise from recent developments, namely:

a) the emergence of the time dependent Schr\"{o}dinger equation from the
time independent Schr\"{o}dinger describing the (approximately) closed\
system of a microscopic quantum element entangled with a classical
environment. Indeed the corresponding TDSE for the quantum system arises in
a disentangling approximation where \textquotedblleft the motion of the
environment provides a time derivative which monitors the development of the
quantum system\textquotedblright . Consequently \textquotedblleft time
enters quantum mechanics only when an external force on the quantum system
is considered classically\textquotedblright \cite{Briggs2,Briggs3,Briggs},
which brings in the laboratory time to the TDSE\cite{Einstein,Moreva,Moreva2}%
.

b) the demise of Pauli's objection that follows from the canonical
quantization of Special Relativity, together with Born's reciprocity
principle\footnote{%
"There is strong formal evidence for the hypothesis, which I have called 
\textit{the principte of reciprocity}, that the laws of nature are
symmetrical with regard to space-time and momentum-energy,or more precisely,
that they are invariant under the transformation \ $\hat{x}_{k}\Rightarrow 
\hat{p}_{k}$ \ ; $\ \hat{p}_{k}\Rightarrow -\hat{x}_{k}$. The most obvious
indications are these: The canonical equations of classical mechanics $\dot{x%
}^{k}=\partial H/\partial p^{k}$ ; \ $\dot{p}^{k}=-\partial H/\partial x^{k}$
are indeed invariant under the transformation, if only the first 3
components of the 4-vectors $x^{k}$ \ and $p^{k}$ \ are considered. These
equations hold also in the matrix or operator form of quantum mechanics.The
commutation rules $[\hat{x}^{k},\hat{p}_{l}]=i\hbar \delta _{l}^{k}$ and the
components of the angular momentum, $m_{kl}=x_{k}p_{l}-p_{l}x_{k},$ show the
same invariance, for all 4 components"\cite{Born}. Although Born did not
succeed in his intended application, the reciprocity principle is currently
subject of renewed interest\cite{Freidel,Morgan}.}\cite%
{Born,Born2,Freidel,Morgan,Morgan2}, as this yields both the Dirac
Hamiltonian $\hat{H}_{D}:=c\mathbf{\alpha .\hat{p}}+\beta m_{0}c^{2}$ and a
self adjoint "time" operator $\hat{T}:=\mathbf{\alpha .\hat{r}}/c+\beta \tau
_{0}$, that restores to position and time the equivalent footing accorded to
energy and momentum in Relativistic Quantum Mechanics (RQM). It introduces
an intrinsic time property $\tau _{0}$ of the system, provides a basis (and
a representation) additional to the usual configuration, momentum and energy
basis and circumvents or really invalidates Pauli's objection\cite%
{Aguillon,Bauer1,Bauer2,Bauer0}. Furthermore this observable complies with
the requirements to condition the other observables in the conditional
interpretation of QG\cite{Bauer10}.

After an abbreviated presentation of such developments, it is finally noted
the need to introduce this new perspective in the extensive developments to
date of both QM and QG where the non existence of a time operator seems to
have been taken for granted, explicit or implicitly.

\section{The emergence of the TDSE\protect\cite{Briggs2,Briggs3,Briggs}}

The Hamilton-Jacobi formulation of classical mechanics is based on assuming
that there is a (sufficient) isolation of the system. Canonical quantization
results in the constraint equation:%
\begin{equation}
\hat{H}\left\vert \Psi \right\rangle =0
\end{equation}%
Introducing the free motion Hamiltonian, Briggs and Rost have shown that the
TDSE emerges from the TISE as a clasical-quantum equation describing the
relational evolution of a quantum system\ $\ S$ \ entangled with an
environment $\ E$ \ that is large enough to be treated classically. Quantum
system \ and environment together are assumed to constitute a (sufficiently)
closed system (the only one completely closed is the universe) to apply a
Hamilton-Jacobi formulation. The wave function satisfying the TISE is
written as:%
\begin{equation}
\Psi (x,R)=\chi _{_{E}}(R)\psi _{_{S}}(x,R).
\end{equation}%
where\ $\{x\}$ and $\{R\}$ refer to the quantum system and environment
coordinates respectively. Projection with the wave functions $\psi
_{_{S}}(x,R)$\ and\ $\chi _{_{E}}(R)$\ \ leads to coupled TISE for both
system and environment. Disentangling approximations are based on assuming
the influence of the environment on the large system is negligeable so that
the environment wave function can be taken as%
\begin{equation}
\chi _{_{E}}(R)=A(R)exp(iW/\hbar )
\end{equation}%
where $W(R,E)$ is the (time-independent) action of the classical free motion
Hamiltonian $H_{E}$. The leading coupling term in the system equation is of
the form:%
\begin{equation}
\frac{1}{M}\frac{\partial W}{\partial R}\frac{\partial }{\partial R}=\frac{P%
}{M}\frac{\partial }{\partial R}=\frac{\partial R}{\partial t}\frac{\partial 
}{\partial R}=\frac{\partial }{\partial t}
\end{equation}%
This introduces the classical time dependence and transforms the TISE for
the quantum system into a TDSE, where the time dependence is assigned to the
state vector - the Schr\"{o}dinger picture. The alternate Heisenberg picture
arises from a unitary transformation that translates the time dependence to
the operators describing the dynamical variables.

The extensive experimental confirmation and applications of the TDSE clearly
identify this time as the laboratory time, thus part of the the spacetime
frame of reference associated to an observer that uses "clocks and measuring
rods"\cite{Einstein}.The classical environment monitors a time evolution of
the atomic system within the full static system. This mechanism has allready
been demonstrated experimentally\cite{Moreva,Moreva2}.

-----------------------------------------------------------------------------------

\textbf{The standpoint is adopted that all time occurring indynamics is
relative in that, if a clock is used to measure time, one is quantifying
positional changes of an observed object by comparison with standard
positional changes of a generalized clock pointer. A closed composite of
several parts is timeless and its states in phase space are described by the
classical TIHJE or the quantum TISE. Observation of one part (the system) by
another (the environment) is an invasive action requiring interaction.
Separating the total action in an \textquotedblleft
adiabatic\textquotedblright\ form allows an approximate time-dependent
dynamics to emerge in which the environment acts as a clock. To function as
a clock the interaction with the system must be negligible.}

-----------------------------------------------------------------------

As this is an extremely schematic presentation, the reader is referred to
the original papers to follow the details of the approximations and their
interesting discussion of the relation to QG.

\section{Time operator in RQM: the demise of Pauli's objection\protect\cite%
{Aguillon}}

The constancy of the speed of light in vacuum for all inertial observers,
denoted by $c$, introduces scalar invariant constraints in time and space,
and in energy and momentum as follows: 
\begin{equation}
p_{\mu }p^{\mu }=\eta ^{\mu \nu }p_{\mu }p_{\nu }=p_{0}^{2}-\mathbf{p}%
^{2}=\pi ^{2}\text{ \ ; \ }x_{\mu }x^{\mu }=\eta ^{\mu \nu }x_{\mu }x_{\nu
}=x_{0}^{2}-\mathbf{r}^{2}=\sigma ^{2}
\end{equation}%
where $\eta ^{\mu \nu }=diag(1,-1,-1,-1).$ Using $c$, the invariants are
defined as $\pi :=m_{0}c$. and similarly $\sigma :=\tau _{0}c$. As $m_{0}$
is identified with the rest mass in any rest frame independent of location, $%
\tau _{0}$ would be also an intrinsic property with dimension of time
evident when the particle is found at the origin $\mathbf{r}=0$ of any
inertial frame independent of its velocity.

The canonical quantization consists in substituting the Hamilton-Jacobi
dynamical variables by self adjoint operators acting on normalized state
vectors in a Hilbert space representing the system and subject to
constraints as follows:%
\begin{equation}
\{\hat{p}_{\mu }\hat{p}^{\mu }-(m_{0}c)^{2}\}\left\vert \Psi \right\rangle =0%
\text{ \ \ \ \ \ ; \ \ \ \ }\{\hat{x}_{\mu }\hat{x}^{\mu }-(\tau
_{0}c)^{2}\}\left\vert \Psi \right\rangle =0
\end{equation}

Factorization (e.g., $\{\hat{p}_{\mu }\hat{p}^{\mu }-(m_{0}c)^{2}\}=(\hat{p}%
_{\mu }+m_{0}c)(\hat{p}^{\mu }-m_{0}c)$), the constraints are satisfied by
the linear equations:%
\begin{equation}
\lbrack \rho ^{\nu }\hat{p}_{\nu }-m_{0}c]\left\vert \Psi \right\rangle =0%
\text{ \ \ \ \ \ \ \ \ ;\ \ \ \ \ \ \ \ \ \ }[\rho ^{\nu }\hat{x}_{\nu
}-\tau _{0}c\left\vert \Psi \right\rangle =0
\end{equation}%
provided that:%
\begin{equation}
\lbrack \hat{p}_{\mu },\hat{p}_{\nu }]=0\text{ \ \ \ ;\ \ \ \ }\{\rho ^{\mu
}\rho ^{\nu }+\rho ^{\nu }\rho ^{\mu }\}=2\eta ^{\mu \nu }\mathbf{I}_{4}%
\text{\ \ \ ;\ \ \ \ \ }[\hat{x}_{\mu },\hat{x}_{\nu }]=0
\end{equation}%
where $\mathbf{I}_{4}$\ is the $4\times 4$. identity matrix. Thus\ the
coefficients $\ \rho ^{\mu }$\ \ obey a Clifford algebra and are represented
by matrices of dimesion four at least\cite{Dirac,Thaller,Schwabl}.

Multiplying by $c\rho ^{0}$\ and defining $\ \rho ^{0}:=\beta ,$ $\ \rho
^{0}\rho ^{i}:=\alpha ^{i}$ one obtains :%
\begin{equation}
c\hat{p}_{0}\left\vert \Psi \right\rangle =\{c\mathbf{\alpha .\hat{p}}+\beta
m_{0}c^{2}\}\left\vert \Psi \right\rangle \text{ \ \ ;\ \ \ }(\hat{x}%
_{0}/c)\left\vert \Psi \right\rangle =\{\mathbf{\alpha .\hat{r}}/c+\beta
\tau _{0}\}\left\vert \Psi \right\rangle
\end{equation}%
that exhibits the Dirac Hamiltonian $\hat{H}_{D}:=c\mathbf{\alpha .\hat{p}}%
+\beta m_{0}c^{2}$ (one recognizes here the procedure followed by Dirac to
obtain a first order linear equation in energy and momentum that agrees with
the second order one resulting from the energy momentum constraint) and in
the same way the time operator $\hat{T}:=\mathbf{\alpha .\hat{r}}/c+\beta
\tau _{0}$, introduced earlier by analogy to the Dirac Hamiltonian\cite%
{Bauer1}.

Finally, the antisymmetrized scalar product of the time space and energy
momentum quadrivectors \ $O^{-}=\eta _{\mu \nu }[x_{\mu },p^{\nu }]$ is the
Lorentz invariant that satisfies Born's reciprocity principle. Upon
quantization it introduces the constraint:%
\begin{equation}
\hat{O}^{-}\left\vert \Psi \right\rangle =\eta _{\mu \nu }[\hat{x}_{\mu },%
\hat{p}^{\nu }]\left\vert \Psi \right\rangle =\{[\hat{x}_{0},\hat{p}%
_{0}]-\delta _{ij}[x_{i}\mathbf{,}p_{j}]\}\left\vert \Psi \right\rangle =0
\end{equation}%
As $(\hat{O}^{-})^{\dagger }=-\hat{O}^{-},$ it is a purely imaginary
constant. The constraint can then be satisfied by the Planck constant $h$
multiplied by any imaginary constant. One can choose:%
\begin{equation}
\lbrack \hat{x}_{0},\hat{p}_{0}]=i3\hbar \text{ \ \ \ \ \ \ \ \ \ \ \ \ \ \ }%
[x_{i}\mathbf{,}p_{j}]=i\hbar \delta _{ij}\text{\ }
\end{equation}%
where $\hbar $\ \ is the reduced Planck constant. This completes the
commutation relations between space and momentum operators from which one
derives:\ a) the $\{x_{i}\}$ and $\{p_{i}\}$ basis are continous from $%
-\infty $\ \ to $+\infty $\ ; b) the representations of the space and
momentum operators in these basis; c) the reciprocal Fourier transform
relation between the state vector representations; and d) the existence of
the Heisenberg uncertainty relation between space and momentum expectation
values. This unified derivation is unfortunately absent in most QM textbooks
where some of the elements are presented as independent \textit{anzats}.

As is well known, the negative energy states are interpreted as representing
antiparticles with positive energy. In the same way the negative time states
would correspond to antiparticles associated with positive times, in
agreement with Feyman's interpretation. Most remarquable, the over the years
much debated question of the introduction of complex numbers in QM, is here
a derived consequence of Born's reciprocity principle\cite{Baylis1,Karam}.

What about Pauli's objection? The time operator being self adjoint is the
generator of a unitary transformation (Stone's theorem\cite{Stone}):%
\begin{equation}
U_{T}=\exp (-i\delta \varepsilon \hat{T}/\hbar )\approx \exp (-i\delta
\varepsilon \mathbf{\alpha .\hat{r}}/c\hbar )\exp (-i\delta \varepsilon
\beta \tau _{0}/\hbar )
\end{equation}%
where the infinitesimal \ $\delta \varepsilon $\ \ is real. In the momentum
representation where $\mathbf{\hat{r}}\Rightarrow i\hbar \nabla _{p}$, this
yields a momentum displacement $\delta \mathbf{p}=\mathbf{(}\delta
\varepsilon \mathbf{/}c^{2}\mathbf{)}c\mathbf{\alpha }=\mathbf{(}\delta
\varepsilon \mathbf{/}c^{2}\mathbf{)}d\mathbf{\hat{r}}/dt$ equal to a
relativistic mass increase multiplied by the relativistic velocity (for a
wave packet $\left\langle d\mathbf{\hat{r}}/dt\right\rangle =v_{gp}$, the
group velovity\cite{Thaller}), and a phase shift $\delta \phi =\beta \delta
\varepsilon \tau _{0}/\hbar .$ Thus $\hat{T}$ acts on the momentum space
where there is no discontinuity nor discrete values, and not on the energy
space as assumed in Pauli' argument, but the change in momentum entails a
change in energy in both energy branches, as $E(\mathbf{p})$\textit{\ }$%
\Rightarrow E(\mathbf{p}+m\mathbf{v}_{gp})$ , circumventing Pauli%
%TCIMACRO{\U{b4}}%
%BeginExpansion
\'{}%
%EndExpansion
s objection.

Equivalently the Dirac Hamiltonian is the generator of a unitary
transformation 
\begin{equation}
U_{H_{D}}=\exp (i\delta t\hat{H}_{D}/\hbar )\approx \exp (i\delta tc\mathbf{%
\alpha .\hat{p}}/\hbar )\exp (i\delta t\beta m_{0}c^{2}/\hbar )
\end{equation}%
for $\delta t$ an infinitesimal time displacement. In the configuration
representation where $\mathbf{\hat{p}}\Rightarrow -i\hbar \nabla _{r},$ this
yields a position displacement $\mathbf{\delta r}=(\delta t)c\mathbf{\alpha
=(}\delta t)d\mathbf{\hat{r}}/dt$ and a phase shift $\delta \phi =\beta
(\delta t)m_{0}c^{2}/\hbar ).$ For a wave packet, $\Psi (\mathbf{r}%
)\Rightarrow \Psi (\mathbf{r}+\mathbf{v}_{gp}\delta t)$.

The eigenvalues $\tau (r)=\pm \sqrt{(r/c)^{2}+\tau _{0}}$ of the time
operator $\hat{T}:=\mathbf{\alpha .\hat{r}}/c+\beta \tau _{0}$ exhibit a
positive and a negative time branches separated by a $2\tau _{0}$ gap in the
same way that the energy spectrum $e(p)=\pm \sqrt{(cp)^{2}+m_{0}c^{2}}$ of $%
\hat{H}_{D}:=c\mathbf{\alpha .\hat{p}}+\beta m_{0}c^{2}$ is composed by a
positive and a negative energy branches separated by a $2m_{0}c^{2}$ gap. As 
$\tau _{0}=h/m_{0}c^{2}$, the de Broglie period\cite{Bauer3}, these gaps are
complementary: to a small energy gap corresponds a large time gap, and
viceversa.

The Heisenberg picture relates the time operator eigenvalues to the
laboratory time as:%
\begin{equation}
\frac{d\hat{T}}{dt}=\frac{1}{i\hbar }\left[ \hat{T},\hat{H}_{D}\right]
=\{I+2\beta K\}+\frac{2\beta }{i\hbar }\{\tau _{0}\hat{H}_{D}-m_{0}c^{2}\hat{%
T}\}
\end{equation}%
where $K=\beta (2\mathbf{s.l}/\hbar ^{2}+1)$ is a constant of motion\cite%
{Thaller}. Integration yields an oscillation (Zitterbewegung) about a linear 
$t$-dependence, as occurs with $\mathbf{\hat{r}}(t)$\cite{Dirac,Thaller}.

To be noted finally is that the association of an intrinsic time property
with mass has been allready exhibited experimentally\cite{Lan}.

\section{The time operator and the conditional interpretation of time in QG%
\protect\cite{Page,Bauer10}}

The Wheeler-deWitt (WdW) equation results from the canonical quantization of
Einstein's GR equations for the Universe, the only truly closed system.
Notwithstanding that the procedure involves a foliation of spacetime into
three-dimension spacelike hypersurfaces and a one-dimension timelike vector
that may characterize the foliation, no time variable appears in the
equation \ The problem of time is that there is no time, leaving indefinite
the choice of foliation among other problems\cite{Anderson}. The WdW
equation predicts a static universe.

Page and Wooters\cite{Page,Kushar2}\textit{\ }advanced the idea that
evolution from the point of view of an internal observer can be introduced
through conditional probabilities between two of the system observables, the
continuum spectrum of one of them serving as the "internal time parameter"
for the other. The conditional probability between projection operators $%
\hat{B}$ and $\hat{C}$ that the observation of $\hat{B}$ is subject to the
observation of $\hat{C}$ is given by:%
\begin{equation}
P(B\mid C)=\frac{\left\langle \Psi \mid \hat{C}\hat{B}\hat{C}\mid \Psi
\right\rangle }{\left\langle \Psi \mid \hat{C}\mid \Psi \right\rangle }
\end{equation}%
\bigskip (where $\left\vert \Psi \right\rangle $ is a solution of\ the
constraint $\hat{H}\left\vert \Psi \right\rangle =0$). The operator $\hat{C}$%
\ must represent an observable with a continuum spectrum, and must not
commute with $\hat{H}$\ as otherwise it would be a constant of motion\cite%
{Kushar2}. It is clear that the intrinsic time operator $\hat{T}$ introduced
satisfies all the necessary requirements, namely:

i)\ it is self adjoint and can represent an observable;

ii) it is a timelike operator, as it is given in terms of the position
operator $\mathbf{\hat{r}}(t)$ whose expectation value represents the
worldline of the system;

iii) it does not commute with the Hamiltonian (see Eq.14) , so it evolves
with time . In the Heisenberg picture one obtains:%
\begin{equation}
\hat{T}(t)=\int_{0}^{t}(1/i\hbar )[\hat{T},\hat{H}_{D}]=\hat{T}%
(0)+\{1+2\beta \hat{K}_{D}\}t+oscillating\ terms
\end{equation}%
where the oscillation occurs about a linear dependence with the laboratory
time, and is present when both positive and negative intrinsic time
components are found in the wave packet. This is in entire analogy with the
evolution of the position operator\cite{Dirac,Thaller}, namely:%
\begin{equation}
\mathbf{\hat{r}(}t\mathbf{)=\hat{r}(}0\mathbf{)+}\{c^{2}\mathbf{\hat{p}/}%
H_{D}\}t+oscillating\ terms(Zitterbewegung)
\end{equation}

iv) its spectrum is a single valued continuum as a function of $t$ in either
positive or negative time branch;

v) it follows that one can construct a free wave packet \ $\left\vert
\varphi _{_{0}}\right\rangle =\int dt_{\nu }c(t_{\nu })\left\vert t_{\nu
}\right\rangle $\ such that $\left\langle \varphi _{_{0}}\left\vert \hat{T}%
\right\vert \varphi _{_{0}}\right\rangle =T(0)$. One can then set $\hat{C}%
_{0}=\left\vert \varphi _{0}\right\rangle \left\langle \varphi
_{0}\right\vert $ and $\hat{B}=\left\vert b\right\rangle \left\langle
b\right\vert $ \ for any operator that represents an observable and commutes
with $\hat{C}$. Then Eq.10\ yields:

\begin{equation}
P\{t_{\nu },b\}\mathbf{=}\frac{\left\vert \Psi (t_{\nu },b)\right\vert ^{2}}{%
\dsum\limits_{\nu }\dint \dint dt_{\nu }^{\prime }db\left\vert \Psi (t_{\nu
}^{\prime },b)\right\vert ^{2}}
\end{equation}%
Consequently $P(t_{\nu },b)$ is interpreted as the joint probability of
finding the system with spin $\nu $\ , eigenvalue $b$ and intrinsic time $%
t_{\nu }=\pm \sqrt{(r/c)^{2}+\tau _{0}^{2}}$.

Contrary to the reservations raised by Unruh and Wald\cite{Unruh}, based
mainly on the now invalidated Pauli's objection, it can be asserted that
this time operator constitutes the additional observable that provides a
time reference by conditionning the other dynamical variables.

\section{Conclusion and Outlook}

Two times are involved in the problem of time in QM and QG, the time present
in the TDSE and an intrinsic time property associated with the system's
mass, which is an eigenvalue of a self adjoint time operator that restores
to time and space that equal footing accorded by SR. They are not unrelated,
as shown by Eq.12, providing an integrated perspective. The continued
experimental confirmation of the predictions and applications of the TDSE
support the identification of the first as the time measured by laboratory
clocks\cite{Einstein}. Although the presentation in this paper is at a basic
level, a fundamental point resides in the invalidation of Pauli's objection
to the existence of a time operator canonically conjugate to energy, which
figures explicit or implicitly in most relevant work on time in QM and QG.
The basic difference resides in that the present derived time operator does
indirectly induce energy changes by acting as a generator of momentum
changes.

As for the existence of the intrinsic time property and the representation
associated to the dynamical time operator, some support is found in the
experimental association of time with mass\cite{Lan}, in the elucidation of
de Broglie's assumption\cite{Baylis}, in electron tunneling and channeling%
\cite{Bauer3,Catillon,Osche} and in simulations of the Dirac equation\cite%
{Gerritsma,Bauer4,Kling,LeBlanc,Dreisow}, but additional evidence should
become a research subject. To quote: "Physics does not depend on the choice
of basis, but which is the most convenient choice depends on the physics"%
\cite{Zee}.I

Taking into account that agreement of QG with relativistic quantum field
theory is to be expected locally\cite{Bonder}, to be investigated is its
inclusion in QFT\cite{Bauer11} and in the extensive developments of QG such
as Loop Quantum Gravity and String Theory\cite{Kiefer}; and whether it
contributes to the solution \ of the diverse facets of the problem of time
in QG\cite{Anderson}, e.g., it may have relevance in the reduction of the
multiple foliation problem of the canonical formulation of QG, providing a
specific choice for the time variable\cite{Kushar2}\textit{\ }with its one
to one correspondence with the timelike worldline $\mathbf{r}(t).$

\end{document}